\documentclass[manuscript,screen]{acmart}
\AtBeginDocument{%
  }

\begin{document}

\title{Atelier à la conférence IHM 2025 : RA Permanente}

\author{Maxime Cauz}
\authornote{Ces auteurs sont les organisateurs de cet atelier.}
\email{maxime.cauz@unamur.be}
\orcid{0000-0002-1234-1772}
\affiliation{%
  \institution{Namur Digital Institute (NaDI), University of Namur}
  \city{Namur}
  \country{Belgium}
}

\author{Thibaut Septon}
\authornotemark[1]
\email{thibaut.septon@unamur.be}
\orcid{0000-0003-0106-0817}
\affiliation{%
  \institution{Namur Digital Institute (NaDI), University of Namur}
  \city{Namur}
  \country{Belgium}
}

\author{Elise Hallaert}
\authornotemark[1]
\email{elise.hallaert@unamur.be}
\orcid{0009-0009-7555-7069}
\affiliation{%
  \institution{Namur Digital Institute (NaDI), University of Namur}
  \city{Namur}
  \country{Belgium}
}

\author{Théo Leclercq}
\authornotemark[1]
\email{theo.leclercq@unamur.be}
\orcid{0009-0005-6201-5712}
\affiliation{%
  \institution{Namur Digital Institute (NaDI), University of Namur}
  \city{Namur}
  \country{Belgium}
}

\author{Bruno Dumas}
\authornotemark[1]
\email{bruno.dumas@unamur.be}
\orcid{0000-0001-5302-4303}
\affiliation{%
  \institution{Namur Digital Institute (NaDI), University of Namur}
  \city{Namur}
  \country{Belgium}
}

\author{Charles Bailly}
\email{charles.bailly@immersion.fr}
\orcid{0000-0002-7797-7368}
\affiliation{%
  \institution{Immersion}
  \city{Bordeaux}
  \country{France}
}

\author{Clément Tyminski}
\email{clement.tyminski-pereira@lecnam.net}
\orcid{}
\affiliation{%
  \institution{National Conservatory of Arts and Crafts (CNAM)}
  \city{Paris}
  \country{France}
}

\author{Matias Peraza}
\email{matias.peraza-reyes@enac.fr}
\orcid{}
\affiliation{%
  \institution{National School of Civil Aviation (ENAC)}
  \city{Toulouse}
  \country{France}
}

\author{Sophie Lepreux}
\email{sophie.lepreux@uphf.fr}
\orcid{0000-0002-0582-7993}
\affiliation{%
  \institution{LAMIH UMR CNRS 8201, Université Polytechnique - Hauts de France}
  \city{Valenciennes}
  \country{France}
}

\author{Emmanuel Dubois}
\email{emmanuel.dubois@irit.fr}
\orcid{0009-0008-9986-4031}
\affiliation{%
  \institution{University of Toulouse, IRIT}
  \city{Toulouse}
  \country{France}
}

\renewcommand{\shortauthors}{Atelier à la conférence IHM 2025 : RA Permanente}

\begin{abstract}
\end{abstract}

\maketitle

\section{Introduction}

À mesure que nous nous dirigeons vers une informatique plus omniprésente, le concept de réalité augmentée permanente (PAR) pourrait conduire à une prochaine évolution majeure de la relation entre les humains, l'informatique et le monde. L'expérience d'un monde augmenté en continu peut avoir des avantages comme des conséquences indésirables sur la vie des utilisateurs, et soulève de nombreuses questions dans de multiples domaines. Dans cet atelier, nous avons souhaité rassembler tous les participants à la conférence IHM'25 préoccupés ou enthousiastes à l'idée de discuter de cette thématique. L’objectif était de faire émerger de l’intelligence collective les défis interdisciplinaires qu’il reste à résoudre pour permettre l’implémentation de ces technologies dans la vie commune, mais également d’en définir les garde-fous nécessaires, est-ce que la PAR n'est pas trop techno-enthousiaste ? L’ensemble de ces éléments ont été regroupés par catégories pour définir un ensemble de futurs grands axes de recherche autour de la réalité augmentée permanente.

\section{Organisation}

L’atelier s’est tenu sur une journée dans le cadre de la conférence IHM’25 organisée à Toulouse. Quatre participants étaient présents durant la session matinale, et trois durant la session de l’après-midi. Seuls deux d’entre eux ont pris part à l’ensemble de la journée.

Au cours de la matinée, les participants ont procédé à une présentation individuelle, incluant une description de leurs travaux de recherche ainsi que de leur conception d’un environnement en réalité augmentée permanente. À l’issue de ces présentations, les organisateurs ont exposé trois visions non exclusives de la réalité augmentée permanente, en les mettant en perspective avec les propositions des participants. La session s’est conclue par un court débat.

L’après-midi, les participants ont été réunis autour d’une table commune. Des scénarios issus de situations de la vie quotidienne leur ont été présentés successivement par l’un des organisateurs. À partir de chaque scénario, les participants étaient invités à conceptualiser l’expérience vécue par un persona dans un contexte de réalité augmentée permanente, en identifiant notamment les avantages, les inconvénients et les limites associées. Deux organisateurs consignaient systématiquement les éléments mentionnés. Après environ 1h30 de discussion autour des scénarios, participants et organisateurs ont produit un Waad\footnote{\cite{hartson2025ux}} en utilisant des post-its reprenant les idées émises. L’objectif du Waad était de faire émerger des regroupements conceptuels et, ce faisant, des catégories thématiques.

\section{Visions présentées par les organisateurs}

Trois visions ont été présentées aux participants. Ces trois visions résultent de l’expertise des organisateurs relative à la thématique de l’atelier et s’appuient sur des travaux scientifiques, des productions artistiques, ainsi que des dynamiques d’innovation et de marketing.

\paragraph{Augmentation de l'environnement} Cette vision est illustrée par le travail de Keiichi Matsuda dans son œuvre Hyper Reality \cite{KeiichiMatsuda2016}, ainsi que dans des travaux scientifiques tels que celui de \citet{cauz2023}. Elle correspond à une amplification de notre environnement, dans lequel l’information virtuelle est continue, contextualisée et distribuée entre divers acteurs. Les éléments présents, qu’ils soient physiques ou numériques, relèvent notamment de la décoration, de l’information, de la sécurité, etc. Le virtuel devient indissociable de l’expérience quotidienne, soulevant la question de la possibilité même de vivre en dehors de ce système.

\paragraph{Informatique spatiale} Cette vision est soutenue par de grandes entreprises telles qu’Apple \cite{apple} et Google \cite{android}, ainsi que par des travaux scientifiques comme celui de \citet{lu2021}. Elle repose sur l’augmentation des capacités humaines, proposant des applications similaires à celles du smartphone rendues continuellement accessibles et projetés directement dans le champ visuel de l’utilisateur. Il s’agit d’une approche davantage centrée sur l’individu, sans toutefois exclure la possibilité de partage d’informations. Par ailleurs, ces informations ne sont pas nécessairement situées (i.e., placées à un endroit précis du monde) ou contextualisées (i.e., relatives à l'environnement).

\paragraph{Métavers} Cette vision est promue par Meta \cite{meta}, représentée dans des œuvres cinématographiques telles que Blade Runner, ainsi que par des travaux scientifiques comme celui de \citet{lik2024}. Bien qu’elle s’appuie principalement sur les technologies de réalité virtuelle plutôt que sur celles de réalité augmentée, elle repose sur l’exploration d’environnements virtuels alternatifs. Les utilisateurs naviguent ainsi entre différents mondes, y compris le monde physique. Les règles régissant ces environnements virtuels peuvent être entièrement reconfigurées afin de s’affranchir de celles qui structurent la réalité. L’ensemble de ces mondes constitue un réseau hautement interconnecté, au sein duquel les données des utilisateurs sont partagées, et où chaque dispositif offre, d’une manière ou d’une autre, un accès continu à ce métavers.

Bien entendu, ces perspectives ne s’excluent pas mutuellement. Les approches centrées sur l’augmentation de l’environnement et celles relatives au métavers présentent des interactions étroites. La réalité augmentée permanente, telle qu’illustrée dans Hyper Reality, peut être interprétée comme un point intermédiaire au sein du continuum du métavers. Les dispositifs actuels permettent par ailleurs une transition fluide entre réalité augmentée et réalité virtuelle, et réciproquement. De même, les visions de l’augmentation de l’environnement et de l’informatique spatiale sont compatibles, comme le démontre Hyper Reality où le protagoniste interagit avec des interfaces assimilables à un smartphone virtuel. L’augmentation de l’environnement implique également l’extension de l’espace sensoriel immédiat. Enfin, les concepts du métavers et de l’informatique spatiale sont également conciliables, dans la mesure où le métavers repose sur la création d’un univers dont les lois physiques et perceptives peuvent être redéfinies. Ainsi, rien n’interdit la virtualisation d’un smartphone sous forme de HUD. Il est même plausible que ces perspectives convergent pour faciliter l’accès à certaines fonctionnalités telles que la déconnexion ou le changement d’univers immersif.

Dans la manière dont l’atelier a été présenté aux participants, la perspective du métavers a été délibérément écartée afin de se concentrer sur l’enrichissement du monde réel. Cependant, il demeurait essentiel de rappeler que la réalité augmentée persistante est fondamentalement associée au concept de métavers.

\section{Activité du matin : Visions initiales des participants}

Lors du workshop, les participants ont exprimé des visions riches et multidimensionnelles de la PAR, la concevant comme un ensemble de technologies et de pratiques capables d’étendre la perception humaine, de soutenir l’action, et de faciliter la collaboration. Six axes majeurs se sont dégagés :

\begin{enumerate}
    \item \textbf{Contrôle de l'IoT via la Réalité Augmentée.} Un participant conçoit la PAR comme un dispositif assurant la continuité entre environnements physiques et numériques, fournissant une interface intuitive, en temps réel et contextualisée, qui permet la paramétrisation de l’IoT sans nécessiter l’usage d’un smartphone. La PAR offre la possibilité de rendre visibles les interactions invisibles.
    \item \textbf{Support aux personnes en situation de handicap.} Un participant conçoit la PAR comme un instrument d'inclusion. De telles solutions pourraient accompagner les personnes ayant des limitations sensorielles, motrices ou cognitives, en enrichissant leur capacité d’autonomie et d’interaction avec l’environnement.
    \item \textbf{Amusement et apprentissage chez les enfants.} Un participant relève un intérêt pédagogique et ludique pour les enfants. La PAR est perçue comme un médium capable de stimuler la curiosité, d’encourager l’exploration interactive et de combiner jeu et apprentissage. Le participant imagine des environnements où les enfants manipulent des éléments virtuels ancrés dans le monde réel, facilitant l’apprentissage expérientiel, l’acquisition de compétences et la motivation.
    \item \textbf{Cocréativité via le partage de modèles 3D.} Un participant décrit la PAR comme un catalyseur de créativité collective. Le partage de modèles 3D en environnement augmenté permettrait une coconception synchrone ou asynchrone, où plusieurs utilisateurs visualisent, modifient et assemblent des artefacts numériques dans un espace partagé.
    \item \textbf{Extension du métavers VR.} Un participant conçoit la PAR comme une extension du métavers en réalité virtuelle vers le monde réel, transformant ce dernier, lorsque la PAR est activée, en un environnement virtuel intégré au métavers. Dans cette perspective, il estime que l’utilisateur ne serait plus contraint de maintenir son attention vers l’écran de son téléphone, les informations étant directement superposées à son champ visuel.
    \item \textbf{Augmentation des objets.} Un participant considère que la PAR peut permettre d’enrichir certains objets grâce à l’ajout d’informations supplémentaires ou d’animations plus aisées à produire en environnement virtuel qu’en situation réelle. Ces objets se trouveraient, par exemple, dans des musées.
\end{enumerate}

Ces six axes se rapportent principalement à la vision de l’informatique spatiale, avec, de manière ponctuelle, la mention des deux autres. Cela peut s’expliquer par le fait que l’atelier se focalise sur la RA ainsi que par la communication marketing actuelle, qui privilégie davantage cette perspective plutôt que celle d’une augmentation de l’environnement. À cet égard, les participants ont souligné que la PAR constitue un vecteur commercial majeur porté par de grandes entreprises, mais qu’elle n’est pas nécessairement souhaitable, dans la mesure où elle soulève de nombreuses questions éthiques, environnementales, relatives à la gestion des données ou encore à la souveraineté. Les candidats identifient l’intérêt de la RA pour certaines activités quotidiennes, notamment l’extension de l’espace de travail, mais ne se déclarent pas prêts à une adoption massive telle que présentée dans la vidéo Hyper-Reality de \citet{KeiichiMatsuda2016}. Toutefois, ils indiquent que son acceptation dépendra étroitement de facteurs culturels, d’un effet générationnel — les jeunes générations étant perçues comme plus enclines à l’adopter — ainsi que d’évolutions dans les modes de construction cognitive.

\section{Activité de l'après-midi : Table ronde et Waad}

Suite aux présentations des visions de chacun, l’atelier s’est organisé autour d’une table ronde au cours de laquelle six scénarios illustrant des contextes d’utilisation potentiels de la RA ont été abordés successivement :
\begin{enumerate}
    \item l’apprentissage des élèves à l’école ;
    \item le partage des espaces publics ;
    \item la visite d’un musée ;
    \item la rencontre d’un ami dans la vie quotidienne ;
    \item le télétravail en réalité augmentée ;
    \item la vie dans une maison connectée.
\end{enumerate}

Le but proposé aux participants était de discuter de l'usage potentiel de la RA étant donné un scénario. Pour ce faire, les participants prenaient la parole de manière informelle et évoquaient les différentes applications possibles de la RA, ainsi que les craintes et espoirs que cela leur évoquaient.
Pendant les discussions, 2 des organisateurs prenaient des notes, dans le but de réaliser un diagramme d’affinité (WAAD) pour la deuxième partie de l'après-midi. Suite aux discussions, les notes d'activité ont été réalisées sur base de la synthèse des notes des deux organisateurs.

\subsection{Diagramme d'affinité}
\label{sec:waad}

\begin{figure}
    \centering
    \includegraphics[width=\linewidth]{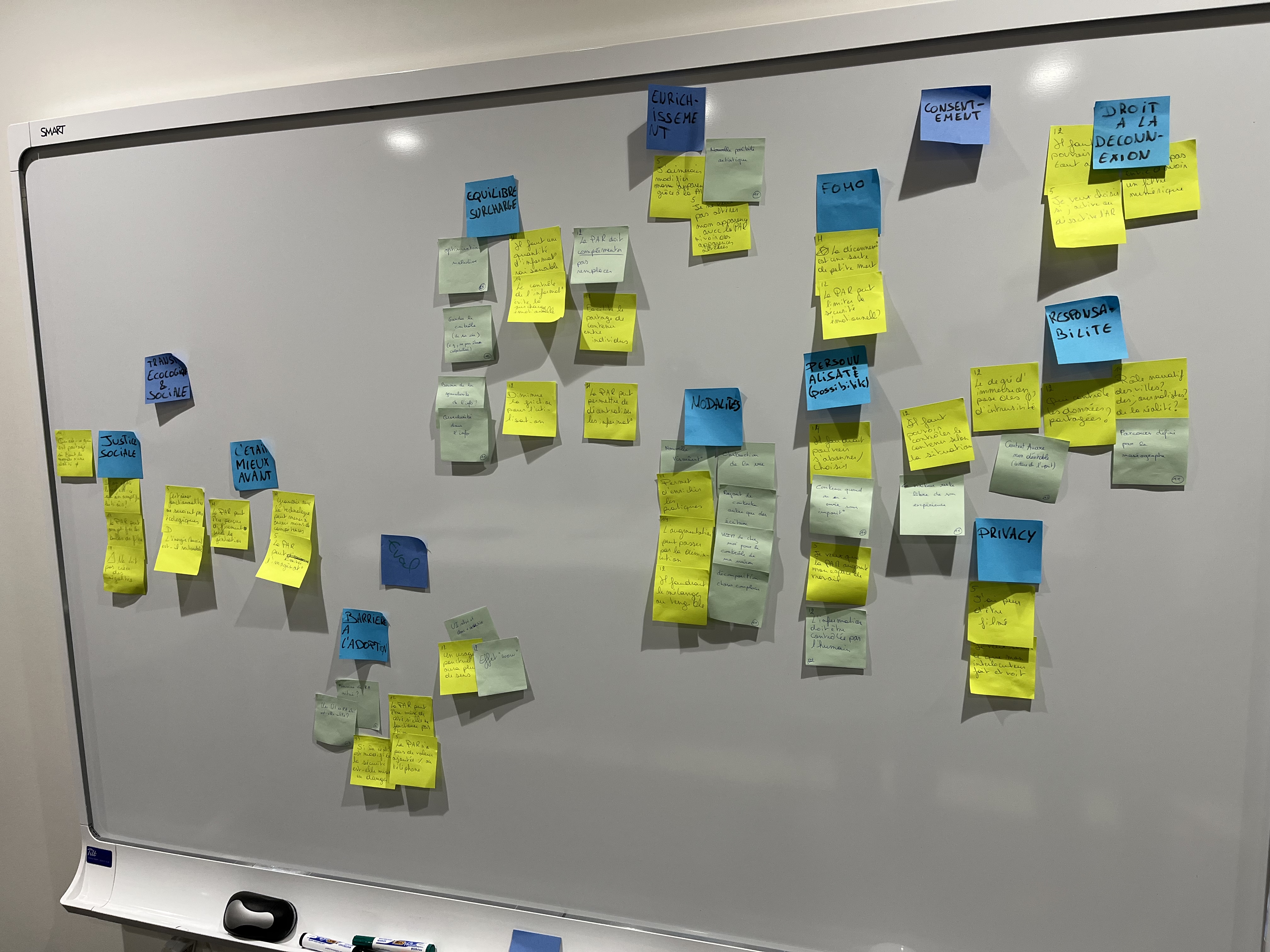}
    \caption{Diagramme d'affinité d'activités de travail (WAAD) réalisé lors du workshop et résumé en section \ref{sec:waad}.}
    \label{fig:waad}
\end{figure}

Plusieurs thématiques ont émergé du WAAD (Figure \ref{fig:waad}), celles-ci sont: 

\begin{enumerate}

    \item \textbf{Transition écologique et sociale.} Cette catégorie a été constituée par deux sous-ensembles. Le premier reprend les enjeux de justice sociale et de développement durable de la RA à grande échelle, et interroge le fondement du besoin d'une telle technologie. Le deuxième souligne la peur des participants que la RA n'isole un peu plus les individus dans des zones de confort ou d'autres formes de chambres d'échos qui leurs seraient propres. Des discriminations, une polarisation des opinions ou une perte de compétences globale pourraient en résulter. 
    
    \item \textbf{Évaluation.} Cette catégorie résulte de la nécessité, selon les participants, de réaliser des évaluations afin de faire adopter la technologie. En conséquence, le sous-titre de cette section, "barrière à l'adoption" nomme l'ensemble d'interrogations que les participants ont vis-à-vis des avantages que peut apporter une RA permanente. Ces barrières recouvrent des questions de sécurité, d'exigences fonctionnelles et de sens.

    \item \textbf{Enrichissement.} La catégorie d'enrichissement a été formée à partir des avantages fonctionnels de la technologie. Une très grande partie des préoccupations qui ont émergés des discussions concernait l'équilibre des informations, c'est-à-dire que la quantité d'informations traitées, visibles ou non, doit être adaptée au contexte, à l'individu, et à une utilisation responsable. Dans un deuxième temps, les modalités d'enrichissement ont été levées : visualisations, obstructions, tangible par exemple. Enfin, l'enrichissement soulignait aussi les disparités parmi les participants en ce qui concerne leurs préférences et volontés en matière de fonctionnalités.
    
    \item \textbf{Consentement.} 
    La plus grande préoccupation concernait le consentement. Celui-ci reprenait des thèmes très précis et interdépendants : le droit à la déconnexion, la personnalisation de ce qui est vu ou encore le "FOMO" (\textit{fear of missing out} ou \textit{peur de rater quelque chose} en français). C'est-à-dire que dans une certaine mesure, le droit à la déconnexion peut être menacé par la peur de se déconnecter, même en présence de la fonctionnalité. 
    Les questions de privacy et de responsabilité ont été citées dans toutes les discussions. De manière générale, les participants craignent de filmer ou d'être filmé sans consentement adéquat, peur de ce qui est fait des données à caractère personnel, veulent que l'utilisateur soit maître de celles-ci, et questionnent la responsabilité des acteurs en jeu: que devient le rôle des villes, des espaces partagés, des acteurs du numérique et des contributeurs au système ?
    
\end{enumerate}

\section*{Remerciements}
Cet atelier a été financé par les projets Wal4XR (n° 2310144) de DigitalWallonia, financés par le Service public de Wallonie (SPW Recherche). Il s'est déroulé dans le cadre de la conférence IHM 2025 qui s'est tenue à Toulouse.

\bibliographystyle{ACM-Reference-Format}
\bibliography{sample-base}

\end{document}